\documentstyle[aps,psfig]{revtex}
\twocolumn
\tighten

\newcommand{\bi}[1]{\bibitem{#1}}

\def\be{\begin{equation}}
\def\ee{\end{equation}}
\def\ba{\begin{eqnarray}}
\def\ea{\end{eqnarray}}
\def\ga{\mathrel{\raise.3ex\hbox{$>$\kern-.75em\lower1ex\hbox{$\sim$}}}}
\def\la{\mathrel{\raise.3ex\hbox{$<$\kern-.75em\lower1ex\hbox{$\sim$}}}}
\newcommand{\fr}[2]{\frac{#1}{#2}}
\begin{document}

\date{\today}

\renewcommand{\thefootnote}{\fnsymbol{footnote}}


\twocolumn[\hsize\textwidth\columnwidth\hsize\csname 
@twocolumnfalse\endcsname

\title{ Electric Dipole Moments in The Limit of Heavy
Superpartners}
\author {Oleg Lebedev and Maxim Pospelov} 
\address{ Centre for Theoretical Physics, University of Sussex, Brighton
BN1 9QJ,~~UK}

\maketitle

\begin{abstract}
Supersymmetric loop corrections induce potentially large 
CP-violating couplings of the Higgs bosons 
to nucleons and electrons that do not vanish in the limit of heavy 
superpartners. The Higgs-mediated CP-odd four-fermion operators 
are enhanced by $\tan^3\beta$ and induce  
electric dipole moments of heavy atoms which 
exceed the current experimental bounds for the electroweak scale 
Higgs masses and $\tan\beta \ga 10$. 
If  only the  first two sfermion
generations are heavy, the Higgs-mediated contributions 
typically dominate over the Barr-Zee type two-loop diagrams 
at $\tan \beta > 30$.
\end{abstract}

\vskip1pc]
 Non--observation of the electric dipole moments (EDMs) of the
neutrons \cite{neutron} and heavy atoms \cite{Tl,Hg,Xe} poses a serious
problem for low--energy supersymmetry. This is because generically the EDMs
are induced already at the one loop level and their predicted values exceed
the experimental limits by orders of magnitude (see
\cite{Abel} for recent analyses).  The most straightforward way to
circumvent this problem is to assume that the superpartner mass  scale  
is rather high (over a few TeV) which leads to the suppression of all
effective one-loop-generated CP-odd operators of dimension five and higher,
and seemingly allows for arbitrary CP-violating phases. 
The arising Higgs mass fine--tuning problem can
be alleviated if the third generation sfermions are  kept light.

In this work, we consider sources of the EDMs which survive
the decoupling of the superpartners while
the Higgs masses  are kept  fixed. In this limit, the Minimal
Supersymmetric Standard Model (MSSM) degenerates into a two-Higgs doublet
model (2HDM) with an important one-loop-induced modification of the Yukawa sector
compared to the usual type II models \cite{thresh}. 
We show that the
Higgs-mediated four-fermion operators induce potentially large electric
dipole moments of heavy atoms that grow as $\tan ^{3}\beta $, and that 
supersymmetric models with large $\tan \beta $ face the SUSY CP problem even
in the limit of heavy SUSY particles. We further compare the
Higgs-mediated contributions with the two-loop induced contributions \cite{W,CKP} 
which  are significant if the masses of the third
generation squarks are  near the electroweak scale.

At the tree level, the down type quarks and charged leptons obtain their
masses from the interaction with the first Higgs doublet $H_{1}$. The finite
one loop SUSY corrections induce considerable couplings of the
second Higgs doublet $H_{2}$ to the $D$-quarks and charged leptons, 
that are absent in the limit of unbroken supersymmetry: 
\ba
-{\cal L}_{Y}&=&Y_{D}^{(0)}H_{1}\bar{D}_{L}D_{R}+{\cal Y}_{D}~H_{2}^{\dagger }\bar{D}%
_{L}D_{R}\nonumber \\
             &+&Y_{E}^{(0)}H_{1}\bar{E}_{L}E_{R}+{\cal Y}_{E}~H_{2}^{\dagger }
\bar{E}_{L}E_{R} +{\rm h.c.}\;,  \label{mass}
\ea
where $Y_{D,E}^{(0)}$ are the tree level Yukawa couplings  and
${\cal Y}_{D,E}$  are the loop induced couplings. The typical
representatives of important threshold diagrams are given in Figs.1a and 1b.
Ignoring  flavor changing effects, we relate $Y_{D,E}^{(0)}$ and  ${\cal Y}_{D,E}$ by: 
\begin{equation}
{\cal Y}_{D}=J_{D}Y_{D}^{(0)},\;\;\;\;\;\;{\cal Y}_{E}=J_{E}Y_{E}^{(0)}\;.
\end{equation}
The loop functions $J_{D}$ and $J_{E}$ are complex and contain the
dependence on the CP phases in the soft-breaking sector, which leads to the
CP-odd interactions of the physical Higgses with the $D$-quarks and leptons.
When one redefines the phase of the right-handed $D$ and $E$ fields, $%
D_{R}\rightarrow e^{-i\delta _{D}}D_{R}$ and $E_{R}\rightarrow e^{-i\delta
_{E}}E_{R}$ such that the mass term in Eq.(\ref{mass}) becomes real, the
induced CP--phase in the Higgs interaction gets enhanced by $\tan \beta
=v_{2}/v_{1}$: $\delta _{D,E}={\rm Arg}(1+J_{D,E}\tan \beta )$. Thus,
large $\tan \beta $ can compensate the loop smallness of $J_{D,E}$ so that
the phases $\delta _{D,E}$ can be order one. An exchange by physical Higgses
will then produce CP--odd four--fermion interactions, Fig.2a. The relevant
interactions are induced by the exchange of a
CP--odd Higgs boson $A$ and CP--even Higgs boson $H$ between the $D$-quarks
and the electron, and between the $D$-quarks: 
\ba
{\cal L}_{4f}\simeq\frac{\tan ^{2}\beta }{2m_{A}^{2}}\sum_{i,j=e,d,s,b}\frac{
Y_{i}^{SM}Y_{j}^{SM}(\sin \delta _{i}-\sin \delta _{j})}{|1+J_{i}\tan \beta
||1+J_{j}\tan \beta |}\nonumber\\\times
\bar{\psi}_{i}\psi_{i}~\bar{\psi}_{j}i\gamma _{5}\psi_{j}\;.
\label{qq}
\ea
Here  $Y_{f}^{SM}$ denote the Standard Model values for the
Yukawa couplings, $Y_{f}^{SM}\equiv \sqrt{2}m_{f}/v$ ($v$=246 GeV),
and the CP--phases are understood modulo $\pi$. In the derivation of 
(\ref{qq}), we have used the relations $m_{A}\simeq m_{H}$, $\cos
^{2}\alpha \simeq 1$, where $\alpha $ is the neutral Higgs mixing angle, and
dropped all $\tan\beta$-suppressed terms. This approximation works
well even for moderately large $\tan \beta $.
\begin{figure}
 \centerline{%
   \psfig{file=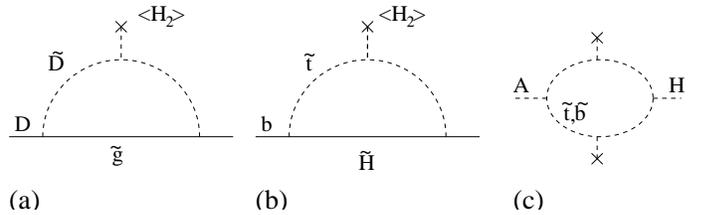,width=9cm,angle=0}%
         }
\vspace{0.1in}
 \caption{$SUSY$ threshold corrections in the down quark Yukawa sector, (a) 
and (b), and in the Higgs sector, (c).}
\end{figure}
CP--odd contact interactions can also be induced via $A-H$ mixing, Fig.2b,
which appears due to CP--violating Higgs couplings to the third generation
squarks \cite{Pilaftsis:1998dd} (Fig.1c): 
\ba
{\cal L}_{AH}\simeq \frac{\langle AH\rangle\tan ^{2}\beta }{2m_{A}^{4}}
\!\!\!\!\!
\sum_{i,j=e,d,s,b}\frac{
Y_{i}^{SM}Y_{j}^{SM}
\bar{\psi}_{i}\psi_{i}~ \bar{\psi}_{j}i\gamma_{5}\psi_{j} }
{|1+J_{i}\tan \beta
||1+J_{j}\tan \beta |},
\label{AH}
\ea
Here we have used $m_{H}^{2}\simeq m_{A}^{2}\gg \langle AH\rangle$. 
Such effects were  studied previously in the context of 2HDMs 
 with spontaneous breaking of CP by Barr \cite{Barr}. 

Inspection of Eq.(\ref{qq}) reveals that the CP-odd coupling grows as $\tan
^{3}\beta $ because $\sin \delta _{i} \simeq $Im$J_{i} \tan \beta /|1+J_{i}\tan \beta
|$, until the radiative corrections become
comparable to the tree-level values. 
Here  we treat $m_A$ as an independent variable (it is proportional to the SUSY 
$B\mu$ parameter).
The cubic growth is different from
a $\tan ^{2}\beta $-behavior in 2HDMs with spontaneous
CP violation \cite{Barr}. Thus, generally  Eq.(\ref{AH}) 
represents a $subleading$ effect, as $\langle AH\rangle$ contains 
a loop smallness  not compensated  by large $\tan\beta$. 
We note that the $QCD$ renormalization group 
flow for the electron--quark interactions 
from $m_{A}$ to $1$GeV is  trivial at one loop and 
 one can simply take $Y_{i}^{SM}$ normalized at 1GeV.
\begin{figure}
 \centerline{%
   \psfig{file=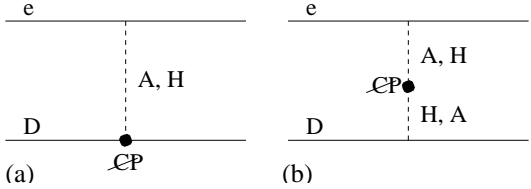,width=7cm,angle=0}%
         }
\vspace{0.1in}
 \caption{Higgs-mediated four-fermion interactions with CP violation 
in the Higgs-fermion vertex (a) and on the Higgs line (b).}
\end{figure}
Using Eq.(\ref{qq}), we calculate the EDMs of paramagnetic
atoms and estimate the EDMs of diamagnetic atoms and neutrons. The semi-leptonic
operators in (\ref{qq}) induce two types of $T$-odd nucleon-electron
interaction 
\begin{equation}
{\cal L}_{CP}=C_{S}\overline{N}N\overline{e}i\gamma _{5}e+C_{P}\overline{N
}i\gamma _{5}N\overline{e}e
\end{equation}
with possible isospin dependence. $C_S$ and $C_P$ are 
severely constrained by the recently improved experimental bounds
on the EDM of the thallium and mercury atoms \cite{Tl,Hg}. 
With the use of the standard
technique for the $QCD$ matrix elements \cite{SVZ} of a heavy quark over a
nucleon state, the isospin-singlet coupling $C_{S}$ can be expressed as 
\ba
C_{S}\simeq\frac{5.5\times 10^{-10}\tan ^{2}\beta }{m_{A}^{2}|1+J_{e}\tan \beta |}
\left[ \frac{(1-0.25\kappa )(\sin \delta _{b}-\sin \delta _{e})}{
|1+J_{b}\tan \beta |}\right. \nonumber\\\left. \!\!\!\!\!
+\frac{3.3\kappa (\sin \delta _{s}-\sin \delta _{e})}{%
|1+J_{s}\tan \beta |}+\frac{0.5(\sin \delta _{d}-\sin \delta _{e})}{
|1+J_{d}\tan \beta |}\right] \;.  
\label{as}
\ea
Here we have  used $(m_{u}+m_{d})\langle N|\overline{u}u+\overline{d
}d|N\rangle /2=45$ MeV and $(m_{u}-m_{d})\langle N|\overline{u}u-\overline{d}%
d|N\rangle /90$ MeV$\ll 1$. The coefficient $\kappa \equiv \langle N|m_{s}%
\overline{s}s|N\rangle /220$ MeV parametrizes the uncertainty in the value of 
$\langle N|m_{s}\overline{s}s|N\rangle $ matrix element. Its ``best''
value $\kappa =1$ is inferred from the leading order flavour $SU(3)$
analysis of the baryon octet mass splittings. An assumption that the strange
quark behaves as a heavy quark would lead to a smaller value, $\kappa =0.3.$
It is important to note that a  significant source of
uncertainty -- the poorly known masses of the light quarks -- does not affect 
Eq.(\ref{as})! 
Using (\ref{as}), and the results of the atomic calculation 
that relates $d_{\rm Tl}$ and $C_{S}$ \cite{KL}, 
\begin{equation}
d_{\rm Tl}\simeq -8.5\times 10^{-17}e~{\rm cm}~ \times C_{S}(100~{\rm GeV})^{2}\;,
\label{dTlCs}
\end{equation}
one can express the thallium EDM in terms of the SUSY parameters. Comparison of 
(\ref{dTlCs}) with the experimental data provides the bound $C_{S}<1.1\times
10^{-8}(100~{ \rm GeV})^{-2}.$

The dimensionless loop functions $J_{i}$'s depend on the pattern of the soft masses. 
To get an idea of the size of the induced EDMs, let us first consider a toy
model with $m_{\mathrm{sfermion}}=m_{\mathrm{gaugino%
}}=|\mu |=|A_{i}|=M\gg M_Z$.  The dominant
contribution comes from the squark-gluino and stop-Higgsino exchange: 
\ba
\label{Jsimple}
&J_e&=0; \;\;\;
J_{d}=J_{s}=\frac{\alpha _{s}}{3\pi }\exp \{i\phi _{\mu}+i\phi_3\}; \\
&J_{b}&=\frac{\alpha _{s}}{3\pi }\exp \{i\phi _{\mu }+i\phi_3\}+
\frac{(Y_{t}^{SM})^{2}}{32\pi ^{2}}\exp \{i\phi _{\mu }+i\phi _{A_t}\},
\nonumber
\ea
where $\phi_\mu, \phi_3, \phi_{A_t}$ are the phases of the $\mu$-parameter, the gluino mass,
 and the $A_{t}$ parameter, respectively.
In the case of general soft terms,
the gluino contribution to $J_i$ should be multiplied by 
$\vert\mu M_3\vert I(m^2_{i1}, m^2_{i2},\vert M_3\vert^2)$ and the Higgsino contribution by
$|\mu A_t|I(m^2_{t1}, m^2_{t2},|\mu|^2)$  \cite{thresh}, 
   where $m_{i1,2}$ are the squark mass eigenvalues,
$M_3$ is the gluino mass, and  
the loop function is defined by
\begin{equation}
I(a,b,c)=2{\frac{ab\ln (a/b)+bc\ln (b/c)+ac\ln (c/a)}{(a-b)(b-c)(a-c)}}\;
\end{equation}
such that $I=1/M^2$ for $a=b=c=M^2$. 
In the same limit, 
the CP-odd Higgs mixing is given by
\ba
\langle AH \rangle = \fr{v^2}{64\pi^2}\left[(Y^{SM}_t)^4
\sin(2\phi_\mu + 2\phi_{A_t}) \right.\nonumber\\
\left.+ (Y^{SM}_b)^4\tan^4\beta\sin(2\phi_\mu + 2\phi_{A_b})\right ].
\ea
Obviously, both $J_i$'s and $\langle AH \rangle$ are independent of 
the superpartner mass scale $M$. An expression for  $\langle AH \rangle$
in a more general case can be found in \cite{Pilaftsis:1998dd}. 
As we will see, the effect of the $A-H$ mixing does not impose significant constraints for 
$m_A \geq 150$ GeV, so henceforth we will mainly 
concentrate on the effect of the vertex corrections.

The Higgs--quark vertex corrections lead to the following 
thallium EDM 
 normalized to  the current 90\% C.L.  experimental bound 
$[d_{\rm Tl}]_{{\rm{\small exp}}}\equiv 9.4\times 10^{-25}{\rm e~cm}$ \cite{Tl}:
\ba
\frac{d_{\rm Tl}}{[d_{\rm Tl}]_{{\rm{\small exp}}}}\simeq \frac{\tan ^{3}\beta }
{350\times m^2_{100}}
\left[ \sin \phi _{\mu }+0.04\sin
(\phi_{\mu }+\phi_{A_t})\right],  \label{Tlsimple}
\ea
where we have set $\phi_3=0$, $|1+J_{i}\tan \beta| \simeq 1$, and $\kappa =1$. $m_{100}$ is 
$m_A$ measured in the units of $100$ GeV. 
Already at $\tan \beta \simeq 7$ the r.h.s. of (\ref
{Tlsimple}) may reach 1 (while the standard EDM contributions are  suppressed 
for the SUSY masses of 10 TeV). 

We conclude that even for arbitrarily heavy 
superpartners,   the SUSY CP problem reappears if
$\tan \beta (m_{A}/{\rm 100 ~GeV)}^{-2/3}\ga 10$. For instance,
with $\phi _{\mu }\sim 1$, $m_{A}\sim $100
GeV, and $\tan \beta \simeq 60$ the induced EDM exceeds the experimental bound by
almost three orders of magnitude!
It is important to note that these calculations are free of large 
nuclear  uncertainties \cite{KL}.

The CP-odd constant $C_{S}$ also induces EDMs of diamagnetic atoms through the
mixing with the hyperfine interaction \cite
{Kozlov,KL}. Our prediction for the mercury EDM is 
\ba
\frac{d_{\rm Hg}(C_{S})}{[d_{\rm Hg}]_{{\rm{\small exp}}}}\simeq
\frac{\tan ^{3}\beta }{900\times m^2_{100}}
\left[ \sin \phi_{\mu }+0.04\sin
(\phi _{\mu }+\phi _{A_t})\right],  \label{HgCs}
\ea
where the current experimental bound is 
$[d_{\rm Hg}]_{{\rm{\small exp}}}\equiv 2\times 10^{-28}{\rm e~ cm}$ \cite{Hg}.
This imposes a slightly weaker bound that the thallium EDM.
However, in the case of diamagnetic atoms there are two additional classes
of contributions, induced by $C_{P}$ and the nuclear Schiff moment. 
To evaluate  $C_{P}$, we follow the strategy of  Ref.\cite{Kolya}.
Unlike the previous case, there is a strong 
dependence of the result on $m_u/m_d$,
and within the error bars for this ratio the matrix elements of 
$\bar s i \gamma_5 s$ and $\bar b i \gamma_5 b$ over the neutron 
are compatible with zero.
The $d$-quark contribution gives
\ba
C_{P}(n)\simeq \frac{6.3\times 10^{-9} }{m_{A}^{2}} 
\frac{\tan ^{2}\beta(\sin \delta _{d}-\sin \delta _{e})}
{|1+J_{e}\tan \beta||1+J_{d}\tan \beta |},
\label{Cp}
\ea
where we have used $m_u/m_d = 0.55$. 
Using the results of the atomic calculation \cite{KL}, we convert this
into the bound
\ba
\frac{d_{\rm Hg}(C_{P})}{[d_{\rm Hg}]_{{\rm{\small exp}}}}\simeq
\frac{\tan ^{3}\beta }{3500\times m^2_{100}}
\sin \phi_{\mu }\;.
  \label{HgCp}
\ea
This is clearly a subleading contribution, compared to (\ref{HgCs}). 
Finally, the $\overline{q}i\gamma _{5}q ~\overline{q}q$ interactions
in (\ref{qq}) induce $d_{\rm Hg}$ via the nuclear Schiff moment.
It is known that a $T$-odd one pion exchange between nucleons is the dominant
source of the Schiff moment. We estimate the $T$-odd pion-nucleon coupling $
\overline{g}_{\pi NN}$ following the approach of \cite{piNN}.  
Since $\sin \delta _{s}=\sin \delta_{d}$ in our case, there are cancellations in the sum
 (\ref{qq}) and the dominant contribution comes from the operator 
$\overline{d}i\gamma _{5}d ~\overline{b}b$.
 The  resulting isospin-triplet coupling $\overline{g}_{\pi NN}\pi^0 \bar NN$ is
\begin{eqnarray}
&&\overline{g}_{\pi NN}\simeq 2.7\times 10^{-13}\frac{\tan ^{2}\beta }{m_{100}^{2}}%
\frac{(\sin \delta _{d}-\sin \delta _{b})(1-0.25\kappa )}{|1+J_{b}\tan \beta
||1+J_{d}\tan \beta |}  \label{pioncoupl}\;. \nonumber
\end{eqnarray}
Skipping a long chain of nuclear and atomic matrix elements that relate $%
\overline{g}_{\pi NN}$ and $d_{\rm Hg}$ (see Ref.\cite{KL} for details), we get:
\begin{equation}
\frac{d_{\rm Hg}({\rm Schiff})}{[d_{\rm Hg}]_{{\rm{\small exp}}}}\simeq\frac{\tan
^{3}\beta }{1.1\times 10^{4}~m^2_{100}} \sin
(\phi _{\mu }+\phi_{A_t}).  
\label{HgSchiff}
\end{equation}
Remarkably, Eq.(\ref{HgSchiff}) has the same 
sensitivity to $\phi _{\mu }+\phi _{A_t}$ as
Eq.(\ref{Tlsimple}), yet this calculation involves considerable uncertainties. 
A combination of 
$d_{\rm Tl}$ and $d_{\rm Hg}$  constrains both phases, 
$\phi _{\mu }$ and $\phi_{A_t}$, once again exemplifying the 
complementarity of the two measurements \cite{FOPR}.

Let us now consider a popular scenario where only the first two sfermion generations 
are assumed to be heavy ($> 10$ TeV). We further assume 
the {\it most conservative} case when the
gluinos are also heavy such that the leading term in Eq.(\ref{Tlsimple}) disappears.
Then we have
\ba
J_{d}=J_{s}=J_{e}=0;\;\;\;J_{b}\simeq \frac{(Y_{t}^{SM})}{32\pi
^{2}}^{2}
A_{t}\mu ~I(m_{\tilde{t}_{1}}^{2},m_{\tilde{t}_{2}}^{2},|\mu |^{2})\;.
\nonumber
\ea
We note, however, that in order  to suppress the contribution of the gluino exchange 
diagram in $J_b$  one would have to require 
$M_3 \ga 60 \mu$!

Fig.3 shows a $\tan\beta$ dependence of 
$C_S/[C_S]_{{\rm{\small exp}}}=d_{\rm Tl}/[d_{\rm Tl}]_{{\rm{\small exp}}}$
in this model.
We use the parameters of Ref.\cite{CKP},
namely  $\vert A_{t,b} \vert=\vert\mu\vert=1$ TeV, Arg$(A_{t,b}\mu)=\pi/2$, 
and $m_{\rm sq}=0.6$ TeV (assuming that the left and right stop and sbottom mass 
parameters are given by $m_{\rm sq}$ while other squarks are decoupled), 
and treat the Higgs masses as $independent$ parameters.
The three curves v1,v2,v3 correspond to $d_{\rm Tl}$ induced
by the Higgs vertex corrections with $m_A=100,200,300$ GeV, 
respectively, while the curve m1 corresponds to 
$d_{\rm Tl}$ induced by the $A-H$ mixing with $m_A=100$ GeV (for $m_A\geq 150$ GeV this
effect does not impose any considerable constraints). 
At $\tan\beta\geq 15$ the generated EDM is comparable to the 
experimental limit and at $\tan\beta\sim 60$ it exceeds the 
experimental limit by up to two orders of magnitude. We observe 
that for this choice of the parameters, the
sensitivity to $\phi_\mu + \phi_{A_t}$ is better than that in Eq.(\ref{Tlsimple})
by a factor of a few.
The sensitivity to $\phi_\mu+\phi_3$ will be as good or better if $M_3 \la 60 \mu$. 
We remark that the considered parameter space is constrained by the observed
$B\rightarrow s\gamma$ branching ratio. When the effect of the SUSY threshold
corrections is taken into account, the constraints become rather weak \cite{Degrassi:2000qf}.

We further compare the Higgs-mediated EDMs with the two-loop 
effects considered in Ref.\cite{CKP}.
The strongest constraint was obtained from the electron or, equivalently,
the thallium atom EDM 
(the neutron EDM imposes weaker constraints, at least in the naive quark model).
At low $\tan\beta$ the two-loop diagrams are more important than 
the ones considered here, while at $\tan\beta \geq 35$ the 
Higgs induced four--fermion operators
provide stronger constraints (see Fig.2 of Ref.\cite{CKP}
subject to the factor of two correction). 
This statement, of course, depends on
$m_A$ and the soft masses. Concerning the $m_A$ dependence,
the EDM due to the Higgs vertex scales as $1/m_A^2$ and
the EDM due to the $A-H$ mixing falls off faster than $1/m_A^4$.
The Barr--Zee contribution scales down roughly linearly with $m_A$ \cite{CKP}
and therefore  is more significant at heavier $m_A$.
On the other hand, the Higgs-mediated EDMs become dominant in the case
of heavier superpartners.

In Fig.4 we present the behavior of different EDM contributions in 
this ``decoupling'' limit, which leads us basically 
to the pattern of SUSY breaking considered before.
In order to observe a cross-over from the two-loop contributions 
to the Higgs mediation in $d_{\rm Tl}$, we fix $\tan\beta=40$ and 
scale the SUSY mass parameters by the common factor $X$: 
$\vert A_{t,b} \vert=\vert\mu\vert=1 X$ TeV, Arg$(A_{t,b}\mu)=\pi/2$, and 
$m_{\rm sq}=0.6 X$ TeV.  
\begin{figure}
 \centerline{%
   \psfig{file=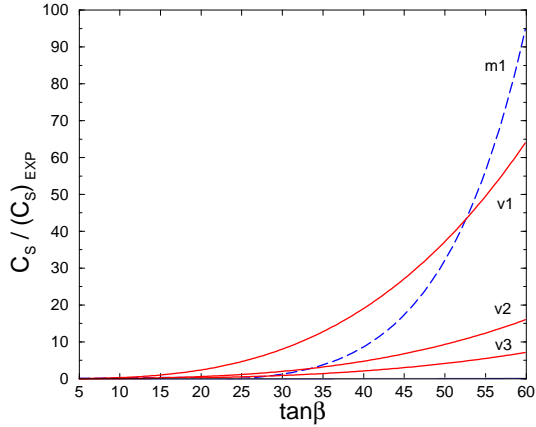,width=7cm,angle=0}%
         }
\vspace{0.1in}
 \caption{The CP--odd electron--nucleon coupling $C_S$ as a function of
$\tan\beta$. The curves v1,v2,v3 correspond to $C_S$ induced
by the Higgs vertex corrections with
$m_A=100,200,300$ GeV. The curve m1 corresponds to $C_S$ 
induced by the $A-H$ mixing with $m_A=100$ GeV. The SUSY parameters are 
$\vert A_{t,b} \vert=\vert\mu\vert=1$ TeV, Arg$(A_{t,b}\mu)=\pi/2$, and 
$m_{{\rm {\small sq}}}=0.6$ TeV.}
\end{figure}
\begin{figure}
 \centerline{%
   \psfig{file=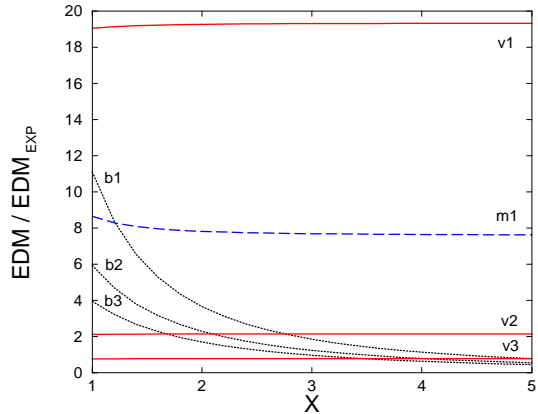,width=7cm,angle=0}%
         }
\vspace{0.1in}
 \caption{ $Tl$ EDM contributions as a function of the SUSY scale factor $X$.
The curves b1,b2,b3 and v1,v2,v3 correspond
to the Barr--Zee type and the CP--odd Higgs vertex contributions to the EDMs
for $m_A=$100,300,500 GeV, respectively. The curve m1 represents
the contribution due to the $A-H$ mixing with $m_A=100$ GeV. The SUSY
parameters are $\tan\beta=40$, $\vert A_{t,b} \vert=\vert\mu\vert=1 X$ TeV, 
Arg$(A_{t,b}\mu)=\pi/2$, and $m_{{\rm {\small sq}}}=0.6 X$ TeV.}
\end{figure}
While the Barr--Zee contributions decouple rather quickly, the effects
we consider stay constant in the limit of heavy superpartners.
Finally, we note that there are the original Barr-Zee contributions 
to the EDMs with the top and bottom quarks in the loop. 
In SUSY models, these are 3-loop effects and do not lead to  
significant constraints (see \cite{Hayashi:1994ha} for 
the analysis in general 2HDMs).

To summarize, we have considered the EDMs of heavy atoms induced by
the  Higgs exchange in SUSY models with  CP violation in the
supersymmetric sector. 
The mechanism of Higgs-mediation is insensitive to an
overall scale of the superpartner masses and grows as $\tan^{3}\beta $ for
fixed values of $m_{A,H}$. It provides significant and 
free of large hadronic and nuclear uncertainties 
constraints on the SUSY CP-phases for  $\tan\beta\ga 10$ and electroweak scale $m_{A,H}$.

\end{document}